# BIN@ERN: ДВОИЧНО-ТРОИЧНОЕ СЖИМАЮЩЕЕ КОДИРОВАНИЕ ДАННЫХ

ИГОРЬ НЕСИОЛОВСКИЙ[1] и АРТЕМ НЕСИОЛОВСКИЙ[2]

Подольск, Московская область, Россия

nesiolovskiy@gmail.com   nesiola@yandex.ru

26 января 2012 г.

Ключевые слова: энтропийное кодирование, сжатие данных без потерь, двоично-троичные префиксные коды

# BIN@ERN: BINARY-TERNARY COMPRESSING DATA CODING

IGOR NESIOLOVSKIY and ARTEM NESIOLOVSKIY

Podolsk, Moscow Region, Russia

January 26, 2012

## АННОТАЦИЯ

Настоящая работа посвящена описанию нового метода экономного кодирования данных, используемых в современных цифровых, компьютерных и телекоммуникационных системах, средствах и устройствах. Он позволяет уменьшать и восстанавливать объем исходных данных, предназначенных для хранения или передачи, без потерь содержащейся в них информации. Метод отличается простотой реализации, высоким быстродействием, плотным сжатием. Он основывается на специальной схеме префиксного кодирования букв алфавита, из которых состоят исходные данные. Эта схема не требует передачи кодовых таблиц букв алфавита от кодера к декодеру, допускает линейные списки представления префиксных кодов букв исходного алфавита, позволяет использовать вычисляемый индекс префиксного кода в линейном списке при декодировании букв, разрешает оценивать степень сжатия до выполнения кодирования, может обходиться без использования операций умножения и деления, вычислений с плавающей точкой при кодировании-декодировании, работоспособна как при статическом (двухпроходном), так и при адаптивном (однопроходном) кодировании, применима к данным с различными по мощности алфавитами, допускает повторное сжатие для его дополнительного уплотнения.

## СОДЕРЖАНИЕ

---

[1] IT-консультант.
[2] Студент-дипломник кафедры кибернетики НИЯУ «МИФИ».

# BIN@ERN: ДВОИЧНО-ТРОИЧНОЕ СЖИМАЮЩЕЕ КОДИРОВАНИЕ ДАННЫХ

## 1. Введение

В современном обществе объемы производимых и используемых данных растут опережающими темпами по сравнению с темпами развития технических средств, применяемых для их хранения и передачи, что обуславливает высокую потребность в технологиях сжатия данных. В создаваемых решениях по сжатию данных без потерь часто применяются такие известные методы энтропийного кодирования данных (и их модификации), как методы Хаффмана (Huffman), Шеннона-Фано (Shannon-Fano), арифметического кодирования (Arithmetic coding). Первые два используют префиксные коды и характеризуются более высокой производительностью (быстродействием), чем третий, но несколько уступают ему по степени сжатия. В них сложнее реализация однопроходного (адаптивного) кодирования исходных данных, но проще реализация двухпроходного (статического) кодирования, чем в третьем методе. Все три метода находят широкое применение в указанной области, поскольку необходимое пользователю решение по сжатию данных это всегда компромисс выбора между степенью сжатия, скоростью кодирования-декодирования, потребляемой оперативной памятью и другими вычислительными ресурсами. При этом они оставляют свободной нишу для быстродействующего метода энтропийного сжимающего кодирования данных без потерь, допускающего простую и эффективную реализацию как двухпроходной, так и однопроходной схем кодирования.

## 2. Некоторые исходные предпосылки

Введем *табличную форму записи* натурального числа $(N)_b$ по целому положительному основанию $b$ с представлением этого числа в бинарном виде, т.е. с использованием нулей и единиц. Она будет состоять из $r$ строк и $c$ столбцов. Строки нумеруются индексом $i = 1 \div r$ сверху вниз, столбцы – индексом $j = 1 \div c$ слева направо. Для определенности будем выражать величины $N$, $b$, $r$, $c$, $i$, $j$ в традиционном десятичном представлении. Для записи числа $(N)_b$ или $(N_{10})_b$ , что то же самое, в табличной форме сначала преобразуем его к поразрядному позиционному представлению вида $a_1 a_2 \ldots a_{c-1} a_c$ по целому основанию $b$, где $a_j = 0 \div (b-1)$ целое числовое значение разряда $j$, а затем значение каждого разряда $a_j$, начиная со старшего (т.е. с первого), зафиксируем в соответствующем столбце табличной формы таким образом, что ячейка таблицы $t_{ij} = 1$ при $i = a_j + 1$, иначе $t_{ij} = 0$. Например, число $(1358_{10})_b$ для $b = 2 \div 5$ будет иметь следующее представление в табличной форме записи:

$(1358_{10})_2 = 10101001110_2$

| 0 | 1 | 0 | 1 | 0 | 1 | 1 | 0 | 0 | 0 | 1 |
|---|---|---|---|---|---|---|---|---|---|---|
| **1** | **0** | **1** | **0** | **1** | **0** | **0** | **1** | **1** | **1** | **0** |

$(1358_{10})_3 = 1212022_3$

| 0 | 0 | 0 | 0 | 1 | 0 | 0 |
|---|---|---|---|---|---|---|
| 1 | 0 | 1 | 0 | 0 | 0 | 0 |
| **0** | **1** | **0** | **1** | **0** | **1** | **1** |

$(1358_{10})_4 = 111032_4$

| 0 | 0 | 0 | 1 | 0 | 0 |
|---|---|---|---|---|---|
| **1** | **1** | **1** | **0** | **0** | **0** |
| 0 | 0 | 0 | 0 | 0 | 1 |
| 0 | 0 | 0 | 0 | 1 | 0 |

$(1358_{10})_5 = 20413_5$

| 0 | 1 | 0 | 0 | 0 |
|---|---|---|---|---|
| 0 | 0 | 0 | 1 | 0 |
| 1 | 0 | 0 | 0 | 0 |
| 0 | 0 | 0 | 0 | 1 |
| **0** | **0** | **1** | **0** | **0** |

Отметим следующие свойства табличной формы записи:

*(2.1):* Количество строк равно значению основания $b$, количество столбцов равно значению $[\log_b(N_{10})]$, округленному с избытком до ближайшего целого.
*(2.2):* Сумма ячеек по любому столбцу записи равна $1$.
*(2.3):* Любую строку табличной формы записи можно восстановить по другим $b - 1$ её строкам, опираясь на свойства *(2.1)* и *(2.2)*.

Назовем строку табличной формы записи *доминантной*, если сумма всех содержащих единицу ячеек в ней не меньше, чем в других строках записи. В приведенном выше примере доминантные строки записей выделены жирным шрифтом. Назовем табличную форму записи *сокращенной*, если из нее исключена любая *недоминантная* строка. Тогда сокращенные формы записи чисел из нашего примера могут принять следующий вид:

$(1358_{10})_2 = 10101001110_2$

| 1 | 0 | 1 | 0 | 1 | 0 | 0 | 1 | 1 | 1 | 0 |
|---|---|---|---|---|---|---|---|---|---|---|

$(1358_{10})_3 = 1212022_3$

| 1 | 0 | 1 | 0 | 0 | 0 | 0 |
|---|---|---|---|---|---|---|
| **0** | **1** | **0** | **1** | **0** | **1** | **1** |

$(1358_{10})_4 = 111032_4$

| **1** | **1** | **1** | **0** | **0** | **0** |
|---|---|---|---|---|---|
| 0 | 0 | 0 | 0 | 0 | 1 |
| 0 | 0 | 0 | 0 | 1 | 0 |

$(1358_{10})_5 = 20413_5$

| 0 | 0 | 0 | 1 | 0 |
|---|---|---|---|---|
| 1 | 0 | 0 | 0 | 0 |
| 0 | 0 | 0 | 0 | 1 |
| **0** | **0** | **1** | **0** | **0** |

Заметим, что сокращенную табличную форму записи числа $(1358_{10})_b$ для $b > 2$ можно привести к еще более экономному битовому представлению, если из каждой ее недоминантной строки исключить избыточные ячейки с нулями, которые могут быть восстановлены по доминантной строке и свойствам *(2.1)* и *(2.2)*, например:

$(1358_{10})_3 = 1212022_3$

| 1 |   | 1 |   | 0 |   |   |
|---|---|---|---|---|---|---|
| **0** | **1** | **0** | **1** | **0** | **1** | **1** |

$(1358_{10})_4 = 111032_4$

| **1** | **1** | **1** | **0** | **0** | **0** |
|---|---|---|---|---|---|
|   |   |   |   | 0 | 1 |
|   |   |   |   | 0 | 1 |

$(1358_{10})_5 = 20413_5$

|   | 0 |   | 1 |   |
|---|---|---|---|---|
| 1 | 0 |   | 0 |   |
| 0 | 0 |   | 0 | 1 |
| **0** | **0** | **1** | **0** | **0** |

Отсюда видно, что в последней форме записи числа $(1358_{10})_b$ для $b > 2$ существует взаимно однозначное соответствие между значениями $b$–ичных разрядов кодируемого числа и образованными в столбцах записи битовыми кодами, начинающимися в ячейках доминантной строки. Замечательно, что образованные битовые коды оказываются *префиксными*, т.е. не являющимися началом других кодов. Следовательно, теперь мы можем получить *экономную форму записи* числа $(1358_{10})_b$ для $b > 2$ в бинарном виде, по которой способны однозначно восстановить исходное десятичное число, зная префиксную кодировку значений $b$–ичных числовых разрядов:

$(1358_{10})_3 = 1212022_3 = 0110110011_{(2)3}$ , где 1 ↔ '01', 2 ↔ '1', 0 ↔ '00'

$(1358_{10})_4 = 111032_4 = 11100000101_{(2)4}$ , где 1 ↔ '1', 0 ↔ '000', 3 ↔ '001', 2 ↔ '01'

$(1358_{10})_5 = 20413_5 = 00100001000101_{(2)5}$ , где 2 ↔ '001', 0 ↔ '0000', 4 ↔ '1', 1 ↔ '0001', 3 ↔ '01'

Отметим, что в нашем примере битовая длина экономной формы записи числа $(1358_{10})_3$ оказалась равна 10 бит, что меньше длины сокращенной табличной формы записи числа $(1358_{10})_b$ при $b \neq 3$.

Зафиксируем следующие выводы:

*(2.4):* Представление числа $(N_{10})_b$ для $b > 2$ в *экономной форме записи* характеризуется наличием в ней *элементарных префиксных кодов*, однозначно соответствующих значениям $b$–ичных разрядов позиционной формы представляемого числа.

*(2.5):* *Элементарные префиксные коды* имеют типовую структуру строения для разных $b > 2$:
'0', '10', '11' для $b = 3$; '0', '10', '110', '111' для $b = 4$; '0', '10', '110', '1110', '1111' для $b = 5$ и т.д.

*(2.6):* Представление числа $(N_{10})_3$ в *экономной форме записи* способно занимать меньший объем двоичной памяти, чем число $N_{10}$, записанное в двоичном виде (или число $(N_{10})_2$).

Теперь оценим *компактность непрерывного представления* первых $N$ натуральных чисел, записанных в *экономной форме* с основанием $b > 2$, по сравнению с их непрерывным представлением в обычном двоичном виде. Непрерывное (потоковое) представление предполагает отсутствие разделителей между числами и их однозначность (распознаваемость). Пусть $N = b^{c_b}$, где $c_b$ – разрядность представления чисел в экономной форме записи с основанием $b$ (или число столбцов в табличной форме записи). Тогда разрядность требуемого двоичного представления $N$ чисел определяется из условия $b^{c_b} = 2^{c_2}$ или $c_2 = \lceil c_b * \log_2 b \rceil$ с избыточным округлением до ближайшего целого. Компактность представления оценивается величиной $\bar{E}_b = c_b/c_2 * \bar{L}_b$, где $\bar{L}_b$ – средняя длина элементарных префиксных кодов в экономной форме записи чисел с основанием $b$. С учетом *(2.5)* справедливо, что $\bar{L}_b = \frac{\left(\sum_1^{b-1} k+(b-1)\right)}{b} = \frac{\left(\sum_1^b k-1\right)}{b} = \frac{\frac{b}{2}*(b+1)-1}{b}$, где $k$ – индекс отдельного элементарного префиксного кода. Тогда $\bar{E}_b = \frac{c_b}{c_2} * \frac{\frac{b}{2}*(b+1)-1}{b} = c_b * \frac{b^2+b-2}{2*b*\lceil c_b*\log_2 b\rceil}$, т.е. полученное соотношение зависит от значений $b$ и $c_b$, выбираемых для рассматриваемой *оценки компактности*. Развивая начатый выше пример экономного представления чисел, оценим $\bar{E}_b$ для $b = 3 \div 8$ и $c_b = 2 \div 12$:

|   |   | $\bar{E}_b$ |   |   |   |   |   |   |   |   |   |
|---|---|---|---|---|---|---|---|---|---|---|---|
| $c_b$ | | 2 | 3 | 4 | 5 | 6 | 7 | 8 | 9 | 10 | 11 | 12 |
| $c_2$ | | 4 | 5 | 7 | 8 | 10 | 12 | 13 | 15 | 16 | 18 | 20 |
| $b$ | 3 | *0,833* | *1,000* | *0,952* | *1,042* | *1,000* | *0,972* | *1,026* | *1,000* | *1,042* | *1,019* | *1,000* |
|   | 4 | 1,125 | 1,125 | 1,125 | 1,125 | 1,125 | 1,125 | 1,125 | 1,125 | 1,125 | 1,125 | 1,125 |
|   | 5 | 1,120 | 1,200 | 1,120 | 1,167 | 1,200 | 1,153 | 1,179 | 1,200 | 1,167 | 1,185 | 1,200 |
|   | 6 | 1,111 | 1,250 | 1,212 | 1,282 | 1,250 | 1,228 | 1,270 | 1,250 | 1,282 | 1,264 | 1,250 |
|   | 7 | 1,286 | 1,286 | 1,286 | 1,286 | 1,361 | 1,350 | 1,342 | 1,335 | 1,330 | 1,369 | 1,361 |
|   | 8 | 1,458 | 1,458 | 1,458 | 1,458 | 1,458 | 1,458 | 1,458 | 1,458 | 1,458 | 1,458 | 1,458 |

Отметим, что оценка $\bar{E}_b = c_b * \frac{b^2+b-2}{2*b*c_b*\log_2 b}$ со снятыми условиями дискретности входящих в нее переменных достигает глобального минимума $\bar{E}_b \approx 0.995$ в точке $b \approx 1.7$. Ближайшей к ней является точка $b = 3$, использованная выше для оценки $\bar{E}_b$.

Сформулируем следующий вывод:

*(2.7):* *Компактность непрерывного представления* первых $N$ натуральных чисел, записанных в *экономной форме* с основанием $b > 2$, по сравнению с их непрерывным представлением в обычном двоичном виде, достигает наилучших значений (в рассмотренных условиях сопоставления) при $b = 3$, т.е. наибольшей компактностью обладает *двоично-троичная* форма экономной записи первых $N$ натуральных чисел.

## 3. Описание метода BIN@ERN

Описанные в разделе 2 данной работы особые свойства троичной системы счисления, проявляемые в двоичной архитектуре хранения данных, задействованы (в качестве одной из отправных точек) при создании метода BIN@ERN. Он является математическим обеспечением автоматизированных процессов сжимающего кодирования исходных данных без потерь, позволяющим уменьшить информационную избыточность в следующих друг за другом серийных элементах (или буквах) исходных данных. Уменьшение избыточности достигается за счет применения специальных *наборов сжимающих префиксных кодов*, заменяющих серийные элементы исходных данных. Результат кодирования данных по этому методу характеризуется меньшими объемами их хранения или передачи, чем требуется для исходных данных. Метод BIN@ERN определяет также и технику декодирования данных, предварительно закодированных в соответствии с ним же. Он может быть реализован на программном или аппаратном уровнях устройств обработки данных, предназначенных для последующего хранения или передачи. Его описание сопровождается рисунками, поясняющими воплощение метода, в которых:

**РИС.1** иллюстрирует схему идентификации наборов сжимающих префиксных кодов метода BIN@ERN;
**РИС.2** иллюстрирует схему формирования сжимающих префиксных кодов в кодовых наборах метода BIN@ERN;
**РИС.3** иллюстрирует схему кодирования, использующую наборы сжимающих префиксных кодов метода BIN@ERN;
**РИС.4** иллюстрирует схему восстанавливающего декодирования, использующую наборы сжимающих префиксных кодов метода BIN@ERN.

### 3.1 Схема идентификации наборов префиксных кодов

Метод сжимающего кодирования данных основывается на использовании специальной схемы префиксного кодирования, из *букв* (или серийных элементов) которого состоят исходные последовательно организованные данные, подлежащие кодированию. Ключевым параметром, который зависит от исходных данных и определяет выбор целесообразного набора префиксных кодов, является истинная мощность $m$ исходного *алфавита* – количество уникальных букв, из которых состоят эти исходные данные. Собственно наборы префиксных кодов детерминированы для различных значений указанного параметра в соответствии с **РИС.1** и идентифицируются порядковым номером $n$, а также границами $M_{min}$ и $M_{max}$, определяющими пределы количества префиксных кодов в каждом кодовом наборе. Причем максимальное количество таких наборов префиксных кодов теоретически не ограничено.

| | *Идентификация наборов префиксных кодов* | |
|---|---|---|
| № набора ($n$) | Минимальное количество кодов в наборе ($M_{min}$) | Максимальное количество кодов в наборе ($M_{max}$) |
| 1 | 3 | 3 |
| 2 | 4 | 9 |
| 3 | 10 | 27 |
| 4 | 28 | 81 |
| 5 | 82 | 243 |
| 6 | 244 | 729 |
| 7 | 730 | 2 187 |
| … … … | … … … | … … … |
| 10 | 19 684 | 59 049 |
| … … … | … … … | … … … |

**РИС. 1**

На **РИС.1** порядковые номера $n$ наборов префиксных кодов определяются последовательными натуральными числами, начиная с 1; минимальное количество префиксных кодов в кодовом наборе с номером $n$ определяется по формуле $M_{min} = 3^{n-1} + 1$; максимальное количество префиксных кодов в кодовом наборе с номером $n$ определяется по формуле $M_{max} = 3^n$. Алфавиты исходных данных мощностью $m = 1$ и $m = 2$ являются вырожденными случаями для рассматриваемого метода, буквы которых элементарно кодируются-декодируются с использованием простейших битовых кодов 0 и 1. Алфавит мощностью $m = 3$ соответствует первому набору префиксных кодов данного метода с идентификатором (порядковым номером) $n = 1$, который отличается от остальных кодовых наборов лишь совпадением границ $M_{min}$ и $M_{max}$, но имеет идентичную с ними схему формирования префиксных кодов.

### 3.2 Схема формирования префиксных кодов в кодовых наборах

Для формирования префиксных кодов, входящих в набор, схема идентификации которого определена в разделе 3.1 данной работы, необходимо выполнить следующие типовые шаги 1-5.
1. Выбрать номер кодового набора $n$ для формирования принадлежащих ему префиксных кодов.
2. Определить базовую разрядную сетку кодирования, в которой число разрядов равно выбранному на шаге 1 номеру $n$ формируемого кодового набора.
3. Сформировать список базовых кодов из символов нулей ('0'), единиц ('1') и двоек ('2'), который
    a. содержит все сочетания '0', '1' и '2' для определенной на шаге 2 и полностью (без пропусков) заполненной разрядной сетки;
    b. упорядочен по убыванию в кодах количества (суммы) нулей ('0');
    c. в группах кодов с одинаковым количеством нулей ('0') дополнительно упорядочен в возрастающем лексикографическом порядке.
4. Для получения результирующего кодового набора в сформированном на шаге 3 списке базовых кодов все единицы ('1') заменить на пары двоичных символов '10', все двойки ('2') заменить на пары двоичных символов '11'.
5. Полученным на шаге 4 префиксным кодам в наборе присвоить порядковые номера последовательными натуральными числами, начиная с 1.

Таким образом, набор префиксных кодов с выбранным номером $n$, схема идентификации которого определена в разделе 3.1, построен. Логика и пример формирования наборов префиксных кодов для кодового набора $n = 3$ проиллюстрированы на **РИС.2**. Максимальное количество базовых кодов в наборе $n$ проверяется по формуле $M_{max} = 3^n$, количество групп кодов $G$ с разным количеством базовых нулей них проверяется по формуле $G = n + 1$, длина префиксных кодов в группе $g \in G$ проверяется по формуле $l_g = 2 * n - z_g$, где $z_g$ равно количеству базовых нулей в группе, количество префиксных кодов в группе $g \in G$ проверяется по формуле $s_g = C_n^{z_g} * 2^{n-z_g}$, где $C_n^{z_g}$ является неупорядоченной выборкой или сочетанием из $n$ элементов по $z_g$ элементов или $C_n^{z_g} = \frac{n!}{z_g! \cdot (n-z_g)!}$ с учетом, что $C_n^n = 1$.

*Пример построения префиксных кодов для кодового набора n = 3*

| Порядковые номера | Базовые коды | Сортированные базовые коды | Количество нулей в группах кодов ($z$) | Результирующие префиксные коды | Длины префиксных кодов ($l$) |
|---|---|---|---|---|---|
| 1 | 000 | 000 | 3 | 000 | 3 |
| 2 | 001 | 001 |   | 0010 |   |
| 3 | 002 | 002 |   | 0011 |   |
| 4 | 010 | 010 | 2 | 0100 | 4 |
| 5 | 011 | 020 |   | 0110 |   |
| 6 | 012 | 100 |   | 1000 |   |
| 7 | 020 | 200 |   | 1100 |   |
| 8 | 021 | 011 |   | 01010 |   |
| 9 | 022 | 012 |   | 01011 |   |
| 10 | 100 | 021 |   | 01110 |   |
| 11 | 101 | 022 |   | 01111 |   |
| 12 | 102 | 101 |   | 10010 |   |
| 13 | 110 | 102 | 1 | 10011 | 5 |
| 14 | 111 | 110 |   | 10100 |   |
| 15 | 112 | 120 |   | 10110 |   |
| 16 | 120 | 201 |   | 11010 |   |
| 17 | 121 | 202 |   | 11011 |   |
| 18 | 122 | 210 |   | 11100 |   |
| 19 | 200 | 220 |   | 11110 |   |
| 20 | 201 | 111 |   | 101010 |   |
| 21 | 202 | 112 |   | 101011 |   |
| 22 | 210 | 121 |   | 101110 |   |
| 23 | 211 | 122 | 0 | 101111 | 6 |
| 24 | 212 | 211 |   | 111010 |   |
| 25 | 220 | 212 |   | 111011 |   |
| 26 | 221 | 221 |   | 111110 |   |
| 27 | 222 | 222 |   | 111111 |   |

**РИС. 2**

### 3.3 Схема сжимающего кодирования

Для выполнения сжимающего кодирования с использованием определенного набора префиксных кодов кодеру необходимо выполнить следующие типовые шаги 1-7.
1. Сформировать алфавит, из букв которого состоят исходные данные.
2. С каждой буквой сформированного на шаге 1 алфавита связать количество ее появлений в исходных данных.
3. Упорядочить полученный на шаге 2 алфавит по убыванию количества появлений его отдельных букв в исходных данных и пронумеровать буквы алфавита последовательными натуральными числами, начиная с 1.
4. Определить мощность алфавита $m$ как число составляющих его уникальных букв и установить по условию $M_{min} \leq m \leq M_{max}$ соответствующий ему порядковый номер $n$ кодового набора, схема идентификации которого определена в разделе 3.1.
5. Получить первые $m$ префиксных кодов для кодового набора с номером $n$, установленным на шаге 4, схема формирования которых определена в разделе 3.2.
6. Присвоить каждой букве алфавита, упорядоченного на шаге 3, начиная с первой и кончая последней, префиксный код, полученный на шаге 5, по условию равенства номера буквы в упорядоченном списке и порядкового номера кода в кодовом наборе.
7. Последовательно от начала к концу заменить в исходных данных буквы алфавита префиксными кодами, присвоенными исходным буквам на шаге 6.
   Таким образом, сжимающее кодирование исходных данных выполнено. Логика и пример сжимающего кодирования для исходных данных, заданных последовательностью букв «ABCDEEFFGGHHHIII», приведены на **РИС.3**.

*Пример выполнения сжимающего кодирования исходных данных, заданных последовательностью "ABCDEEFFGGHHHIII"*

| Порядковые номера | Буквы алфавита | Количество появлений в исходных данных ($k$) | Выбор кодового набора | Префиксные коды из набора $n = 2$ | Длины префиксных кодов ($l$) | Суммарная длина закодированных букв ($k*l$) |
|---|---|---|---|---|---|---|
| 1 | H | 3 | | 00 | 2 | 6 |
| 2 | I | 3 | | 010 | | 9 |
| 3 | E | 2 | Мощность исходного алфавита $m = 9$ | 011 | 3 | 6 |
| 4 | F | 2 | | 100 | | 6 |
| 5 | G | 2 | | 110 | | 6 |
| 6 | A | 1 | Т.к. $4 < m \leq 9$, следовательно $n = 2$ | 1010 | | 4 |
| 7 | B | 1 | | 1011 | 4 | 4 |
| 8 | C | 1 | | 1110 | | 4 |
| 9 | D | 1 | | 1111 | | 4 |
| ИТОГО букв: | | 16 | | | ИТОГО бит: | 49 |

Исходные данные: ABCDEEFFGGHHHIII

Закодированные данные: 1010101111101111011011100100110110000000010010010

**РИС. 3**

### 3.4 Схема восстанавливающего декодирования

Для однозначного восстановления исходных данных, закодированных по схеме раздела 3.3 с использованием определенного набора префиксных кодов, декодеру необходимо выполнить следующие типовые шаги 1-5.
1. Получить от кодера идентификационный номер $n$ использованного при кодировании набора префиксных кодов.
2. Получить от кодера упорядоченный ранее на шаге 3 схемы раздела 3.3 алфавит из $m$ букв, из которого состоят исходные данные, и пронумеровать буквы алфавита последовательными натуральными числами, начиная с 1.
3. Сформировать первые $m$ префиксных кодов для кодового набора с номером $n$, установленным на шаге 1, схема формирования которых определена в разделе 3.2.
4. Присвоить каждому префиксному коду, сформированному на шаге 3, начиная с первого и кончая последним, букву алфавита из числа полученных на шаге 2, по условию равенства порядкового номера кода в кодовом наборе и номера буквы в упорядоченном алфавитном списке.

*Пример выполнения восстанавливающего декодирования для закодированных данных, представленных битовой последовательностью "1010101111101111011011100100110110000000010010010"*

| Полученный номер кодового набора | Полученные буквы исходного алфавита | Порядковые номера букв | Префиксные коды из набора $n = 2$ | Порядковые номера кодов | Выделение префиксных кодов из набора $n = 2$ в закодированных данных и восстановление исходных данных |
|---|---|---|---|---|---|
| | H | 1 | 00 | 1 | |
| | I | 2 | 010 | 2 | |
| | E | 3 | 011 | 3 | 1010'1011'1110'1111'011'011'100'100'110'110'00'00'00'010'010'010 |
| $n = 2$ | F | 4 | 100 | 4 | |
| | G | 5 | 110 | 5 | A'B'C'D'E'E'F'F'G'G'H'H'H'I'I'I |
| | A | 6 | 1010 | 6 | |
| | B | 7 | 1011 | 7 | |
| | C | 8 | 1110 | 8 | |
| | D | 9 | 1111 | 9 | |
| ИТОГО букв: | | 9 | | | |

Закодированные данные: 1010101111101111011011100100110110000000010010010

Восстановленные данные: ABCDEEFFGGHHHIII

**РИС. 4**

5. Последовательно от начала к концу заменить в закодированных данных, сформированных кодером по схеме раздела 3.3, префиксные коды буквами алфавита, присвоенными кодам на шаге 4.

   Таким образом, восстанавливающее декодирование закодированных исходных данных выполнено. Логика и пример восстанавливающего декодирования для закодированных данных, представленных битовой последовательностью «10101011110111101101110010011011000000010010010», приведены на **РИС.4**.

### 3.5  Комментарии по реализации метода

Поскольку используемые в методе BIN@ERN наборы сжимающих префиксных кодов детерминированы, то они могут либо независимо генерироваться кодером и декодером непосредственно в процессе кодирования-декодирования, либо быть заранее сформированы и сохранены в памяти, доступной кодеру и декодеру, а именно: или в памяти устройства кодирования/декодирования, или в локальной/глобальной сети (intranet/internet).

Последовательности битовых символов в префиксных кодах (сигнатуры кодов) и последовательности префиксных кодов в кодовых наборах метода BIN@ERN удобны для определения вычисляемого индекса кодового слова в линейном списке префиксных кодов набора при восстанавливающем декодировании исходных данных и ускорения процесса декодирования. Очевидно, что существуют инварианты следования битов в префиксных кодах (например, если все битовые символы '0' и '1' в них заменить друг на друга взаимоисключающим образом) и следования префиксных кодов в кодовых наборах (например, если в каждой из групп префиксных кодов $g \in G$ их произвольно пересортировать), которые, однако, не могут привести к большему сжатию исходных данных, чем сжатие, основанное на предлагаемой кодировке.

## 4.  Результаты исследования метода BIN@ERN

Для экспериментального исследования сжимающей способности метода BIN@ERN реализована соответствующая компьютерная программа. Она позволяет кодировать и декодировать файлы согласно описанному в настоящей работе методу. Для выполнения кодирования программе задается входной параметр $L$ – битовая длина серийного элемента (буквы) входного файла. В результате кодирования создается выходной файл, в структуре которого выделяются

– постоянная служебная часть (12 байт),
– алфавит, из букв которого состоит исходный файл,
– последовательность закодированных методом BIN@ERN исходных данных входного файла.

Для точного восстанавливающего декодирования исходных данных без потерь программе достаточно данных, содержащихся непосредственно в выходном файле, полученном при кодировании.

Сжимающая способность метода BIN@ERN исследована на стандартном тестовом наборе Canterbury Corpus (http://corpus.canterbury.ac.nz/descriptions/#cantrbry), содержащем 11 исходных файлов, с помощью указанного выше программного приложения. Результаты сжатия зафиксированы в представленной ниже таблице, в которой достигнутая компрессия данных оценивается в битах на байт (бит/байт) и в процентах по следующим соотношениям:

***Компрессия***, бит/байт = (***Длина сжатого файла*** / ***Длина исходного файла***) * **8**

***Компрессия***, % = (***Длина сжатого файла*** / ***Длина исходного файла***) * **100**

| Исходный файл | Исходная длина | Сжатая длина | Компрессия | | Сжатая длина | Компрессия | |
|---|---|---|---|---|---|---|---|
| | *байт* | *байт* | *бит/байт* | *%* | *байт* | *бит/байт* | *%* |
| | | | $L = 8$ | | | $L = 16$ | |
| alice29.txt | 152089 | 99159 | 5,2158 | 65,20 | 88630 | 4,6620 | 58,28 |
| asyoulik.txt | 125179 | 83449 | 5,3331 | 66,66 | 74026 | 4,7309 | 59,14 |
| cp.html | 24603 | 19911 | 6,4743 | 80,93 | 16835 | 5,4741 | 68,43 |
| fields.c | 11150 | 8794 | 6,3096 | 78,87 | 7058 | 5,0640 | 63,30 |
| grammar.lsp | 3721 | 2515 | 5,4071 | 67,59 | *2583* | *5,5533* | *69,42* |
| kennedy.xls | 1029744 | 876754 | 6,8114 | 85,14 | 556918 | 4,3267 | 54,08 |
| lcet10.txt | 426754 | 328823 | 6,1642 | 77,05 | 247828 | 4,6458 | 58,07 |
| plrabn12.txt | 481861 | 315007 | 5,2298 | 65,37 | 275411 | 4,5725 | 57,16 |
| ptt5 | 513216 | 333252 | 5,1947 | 64,93 | 272503 | 4,2478 | 53,10 |

| Исходный файл | Исходная длина | Сжатая длина | Компрессия | | Сжатая длина | Компрессия | |
|---|---|---|---|---|---|---|---|
| | *байт* | *байт* | *бит/байт* | *%* | *байт* | *бит/байт* | *%* |
| | | | $L=8$ | | | $L=16$ | |
| sum | 38240 | 34800 | 7,2803 | 91,00 | 28846 | 6,0347 | 75,43 |
| xargs.1 | 4227 | 2927 | 5,5396 | 69,25 | *3105* | *5,8765* | *73,46* |
| **СУММАРНО:** | **2810784** | **2105391** | **5,9923** | **74,90** | **1573743** | **4,4792** | **55,99** |

Отсюда видно, что метод предполагает возможность как «мягкого», так и «плотного» сжатия, что соответствует известным положениями теории информации. Отдельные эффекты незначительного превышения длины «плотно» сжатых файлов над «мягко» сжатыми, характерные для обработки исходных тестовых файлов малой длины, объясняются ростом их информационной энтропии при смене алфавитов исходных данных.

Для дополнительного уменьшения длины выходных файлов с «мягким» сжатием возможно их повторное сжатие, допускаемое методом BIN@ERN. Примеры однократного дожатия файлов, предварительно сжатых методом BIN@ERN, представлены в следующей таблице:

| Исходный файл | Исходная длина | Сжатая длина | Длина повторного сжатия | | |
|---|---|---|---|---|---|
| | *байт* | *байт* | *байт* | | |
| | | $L=8$ | $L=3$ | $L=6$ | $L=9$ |
| alice29.txt | 152089 | 99159 | 95441 | 94128 | 93216 |
| asyoulik.txt | 125179 | 83449 | 81898 | 80756 | 79925 |
| cp.html | 24603 | 19911 | 19022 | 18804 | *18980* |
| fields.c | 11150 | 8794 | 8107 | 8058 | *8387* |
| grammar.lsp | 3721 | 2515 | 2438 | *2444* | *2798* |
| kennedy.xls | 1029744 | 876754 | 715266 | 711209 | 706497 |
| lcet10.txt | 426754 | 328823 | 298993 | 295819 | 293594 |
| plrabn12.txt | 481861 | 315007 | 303696 | 299038 | 294854 |
| ptt5 | 513216 | 333252 | 237861 | 237352 | 237351 |
| sum | 38240 | 34800 | 30860 | 30593 | *30863* |
| xargs.1 | 4227 | 2927 | 2917 | 2913 | *3281* |
| **СУММАРНО:** | **2810784** | **2105391** | **1796499** | **1781114** | **1769746** |

Полученные результаты свидетельствуют, что дожатие вполне практично, но не беспредельно и предполагает существование определенного оптимума уплотнения результирующих данных.

Для дополнительного уменьшения длины выходных файлов с «плотным» сжатием возможно сжатие алфавитных данных, присутствующих в этих файлах, допускаемое методом BIN@ERN. Примеры сжатия алфавитных данных таких файлов представлены в следующей таблице, последний столбец которой содержит оценку результата (эффекта) сжатия алфавита в % от длины файла, предварительно сжатого методом BIN@ERN:

| Исходный файл | Сжатая длина | В т.ч. длина алфавита | Сжатая длина алфавита | Результат сжатия алфавита |
|---|---|---|---|---|
| | *байт* | *байт* | *байт* | *%* |
| | $L=16$ | | $L=8$ | |
| alice29.txt | 88630 | 2265 | 1768 | 0,56 |
| asyoulik.txt | 74026 | 2087 | 1612 | 0,64 |
| cp.html | 16835 | 2385 | 2107 | 1,65 |
| fields.c | 7058 | 1290 | 1179 | 1,57 |
| grammar.lsp | 2583 | 709 | 587 | 4,72 |
| kennedy.xls | 556918 | 3310 | *3492* | *-0,03* |
| lcet10.txt | 247828 | 3428 | 3038 | 0,16 |
| plrabn12.txt | 275411 | 2177 | 1691 | 0,18 |
| ptt5 | 272503 | 4642 | 4150 | 0,18 |
| sum | 28846 | 4778 | *5011* | *-0,81* |
| xargs.1 | 3105 | 885 | 712 | 5,57 |

# BIN@ERN: ДВОИЧНО-ТРОИЧНОЕ СЖИМАЮЩЕЕ КОДИРОВАНИЕ ДАННЫХ

Сжатие алфавитных данных, как и предыдущую меру дожатия файлов, резонно применять с учетом знания моделей строения этих данных, статистические характеристики которых известны по результатам первичного сжатия исходных данных.

В данной работе не ставилась задача оптимизации быстродействия экспериментальной компьютерной программы, реализующей метод BIN@ERN. Поэтому здесь к ней применяется относительная оценка эффективности функционирования, позволяющая определить экономичность метода при переходе от сжатия файлов с параметром $L=8$ к сжатию файлов с параметром $L=16$, которая уменьшает влияние факторов, связанных со спецификой конкретной программной реализации метода и особенностями конкретной программно-аппаратной среды исполнения экспериментального приложения. Для такой оценки используется следующие соотношения:

*Цена экономии* = *Время сжатия* / (*Исходная длина* – *Сжатая длина*)

*Изменение цены экономии*, % = (*Цена экономии*$_{L=16}$ / *Цена экономии*$_{L=8}$ – **1**) * **100**

| Исходный файл | Исходная длина | Сжатая длина | | Изменение цены экономии | |
|---|---|---|---|---|---|
| | байт | байт | байт | % | тренд |
| | | $L=8$ | $L=16$ | | |
| alice29.txt | 152089 | 99159 | 88630 | 3 | ↗ |
| asyoulik.txt | 125179 | 83449 | 74026 | 6 | ↗ |
| cp.html | 24603 | 19911 | 16835 | 1 | ↗ |
| fields.c | 11150 | 8794 | 7058 | -15 | ↘ |
| grammar.lsp | 3721 | 2515 | *2583* | 25 | ↗ |
| kennedy.xls | 1029744 | 876754 | 556918 | -67 | ↘ |
| lcet10.txt | 426754 | 328823 | 247828 | -34 | ↘ |
| plrabn12.txt | 481861 | 315007 | 275411 | -6 | ↘ |
| ptt5 | 513216 | 333252 | 272503 | -20 | ↘ |
| sum | 38240 | 34800 | 28846 | 1 | ↗ |
| xargs.1 | 4227 | 2927 | *3105* | 77 | ↗ |

Цена экономии объема хранения данных (памяти) это затраты машинного времени на единицу объема высвобождаемого пространства в результате сжатия файла. Рост этой цены может свидетельствовать о снижении экономичности сжатия, снижение цены – об увеличении экономичности. Полученные результаты показывают, что сжатие исходных файлов среднего и большого размера методом BIN@ERN скорее более экономично с параметром $L=16$ (т.е. с большей битовой длиной буквы входного алфавита или с большей его мощностью), чем с параметром $L=8.$

Тестирование метода BIN@ERN в предельных условиях работы, а именно, при обработке несжимаемых данных, представляет преимущественно теоретический, нежели практический интерес. Частично оценка эффективности кодирования подобных данных представлена в таблице раздела 2 настоящей работы в виде значений $\bar{E}_b$ для $b=3$. В следующей таблице эта оценка расширена на результаты двоично-троичного префиксного кодирования (2-3 кодирования) исходных данных, состоящих из алфавитов, мощности которых кратны степеням двойки и буквы которых равновероятны в этих данных.

| Длина двоичного кода | Мощность алфавита | Минимальная длина 2-3 кода | Максимальная длина 2-3 кода | Избыточность кодирования (%) |
|---|---|---|---|---|
| 3 | 8 | 2 | 4 | 8,33 |
| 4 | 16 | 3 | 5 | 12,50 |
| 5 | 32 | 4 | 6 | 13,75 |
| 6 | 64 | 4 | 7 | 5,47 |
| 7 | 128 | 5 | 8 | 7,25 |
| 8 | 256 | 6 | 10 | 9,38 |
| 9 | 512 | 6 | 11 | 5,01 |
| 10 | 1024 | 7 | 12 | 6,01 |
| 11 | 2048 | 7 | 13 | 4,67 |
| 12 | 4096 | 8 | 14 | 4,64 |
| 13 | 8192 | 9 | 15 | 5,29 |
| 14 | 16384 | 9 | 16 | 4,25 |
| 15 | 32768 | 10 | 17 | 4,28 |

| Длина двоичного кода | Мощность алфавита | Минимальная длина 2-3 кода | Максимальная длина 2-3 кода | Избыточность кодирования (%) |
|---|---|---|---|---|
| 16 | 65536 | 11 | 18 | 4,73 |
| 17 | 131072 | 11 | 19 | 3,91 |
| 18 | 262144 | 12 | 20 | 3,95 |
| 19 | 524288 | 12 | *23* | *5,01* |
| 20 | 1048576 | 13 | 22 | 3,63 |

В приведенных результатах отмечается тенденция постепенного уменьшения избыточности кодирования несжимаемых данных методом BIN@ERN. Платой за уменьшение избыточности является увеличение мощности кодируемого алфавита. Это еще раз наводит на вполне практическую мысль о том, что оптимум результата сжатия этим методом обычных данных определяется достижением рациональной пропорции между объемами сжатых данных и данных исходного алфавита.

## 5. Развитие метода BIN@ERN

В настоящей работе рассмотрена реализация статического алгоритма метода BIN@ERN. Представляет интерес разработка его адаптивной версии, а так же возможность их сравнения между собой по показателям эффективности кодирования-декодирования. При разработке целесообразно обратить внимание прежде всего на такие потенциальные преимущества однопроходного метода BIN@ERN, как отсутствие кодового дерева и необходимости его оперативной перестройки, независимость префиксных кодов от исходных данных и их изначальная детерминированность (заданность), возможность оптимизации схемы упорядочивания кодируемых букв алфавита в линейном списке с учетом того, что в пределах группы $g \in G$ двоично-троичные префиксные коды букв имеют одинаковую битовую длину, а значит, не имеют преимущества друг перед другом по стоимости кодирования.

## 6. Заключение

В данной работе изложены базовые математические предпосылки метода BIN@ERN, подробно описан статический алгоритм метода и представлены контрольные примеры его применения, приведены развернутые результаты его исследования, полученные с помощью экспериментального программного приложения, реализующего данный метод. Согласно полученным результатам, он демонстрирует хорошие показатели двоично-троичного префиксного кодирования-декодирования данных, предварительно подтверждающие возможность его эффективного использования в программно-технических решениях по сжатию данных. У метода BIN@ERN также просматривается хорошая перспектива реализации на его базе адаптивного алгоритма кодирования данных без потерь. Вместе с тем очевидно, что метод нуждается в дополнительном анализе с целью подтверждения его полезных потребительских возможностей путем проведения экспериментов по сжатию других тестовых наборов файлов или типовых тестовых данных, сравнения результатов его работы с классическими методами энтропийного кодирования в сопоставимых условиях и режимах функционирования.

При разработке метода BIN@ERN не возникало потребностей в технической литературе по его непосредственной тематике. Но при обработке результатов исследования метода, представленных в настоящей статье, такая литература понадобилась для их итогового сравнения с достижениями других исследователей. Ее поиск в Интернете пока не привел к успеху, но встретилась публикация [1], которая свидетельствует как о проводимых в данном направлении работах, так и о малочисленности соответствующих статей и материалов в открытых источниках со свободно распространяемой информацией. Подходы и результаты, представленные в статье [1] и в настоящей работе, содержательно отличаются, поэтому у читателей имеется возможность более объемно оценить предлагаемый здесь метод BIN@ERN.

## Литература

[1] Udita Katugampola, *A new technique for text data compression,* http://arxiv.org/pdf/1012.4241.pdf, 20 Dec 2010